\documentclass[proc]{edpsmath}
\usepackage[utf8]{inputenc}
\usepackage[T1]{fontenc}
\usepackage{amsmath,amssymb,amsthm}
\usepackage{bbm}
\usepackage{xparse}
\usepackage{color}
\usepackage{bbm}
\usepackage{verbatim}
\usepackage{enumitem}
\setlist[itemize]{leftmargin=*}
\setlist[enumerate]{leftmargin=*,label=(\arabic*),ref=(\arabic*)}
\usepackage[numeric,initials,nobysame,msc-links,abbrev]{amsrefs}
\renewcommand{\eprint}[1]{\href{https://arxiv.org/abs/#1}{arXiv:#1}}
\newcommand{\pageafter}[1]{#1~pp.}
\BibSpec{article}{%
+{} {\PrintAuthors} {author}
+{,} { \textit} {title}
+{.} { } {part}
+{:} { \textit} {subtitle}
+{,} { \PrintContributions} {contribution}
+{.} { \PrintPartials} {partial}
+{,} { } {journal}
+{} { \textbf} {volume}
+{} { \PrintDatePV} {date}
+{,} { \issuetext} {number}
+{,} { \pageafter} {pages}
+{,} { } {status}
+{,} { \PrintDOI} {doi}
+{,} { available at \eprint} {eprint}
+{} { \parenthesize} {language}
+{} { \PrintTranslation} {translation}
+{;} { \PrintReprint} {reprint}
+{.} { } {note}
+{.} {} {transition}
+{} {\SentenceSpace \PrintReviews} {review}
}
\BibSpec{collection.article}{%
+{} {\PrintAuthors} {author}
+{,} { \textit} {title}
+{.} { } {part}
+{:} { \textit} {subtitle}
+{,} { \PrintContributions} {contribution}
+{,} { \PrintConference} {conference}
+{} {\PrintBook} {book}
+{,} { } {booktitle}
+{,} { \PrintDateB} {date}
+{,} { \pageafter} {pages}
+{,} { } {status}
+{,} { \PrintDOI} {doi}
+{,} { available at \eprint} {eprint}
+{} { \parenthesize} {language}
+{} { \PrintTranslation} {translation}
+{;} { \PrintReprint} {reprint}
+{.} { } {note}
+{.} {} {transition}
+{} {\SentenceSpace \PrintReviews} {review}
}

\usepackage[capitalise]{cleveref}

\usepackage{pgf,tikz}
\usetikzlibrary{arrows}
\usetikzlibrary{patterns}

\newtheorem{thm}{Theorem}
\crefname{thm}{Theorem}{Theorems}

\crefname{cor}{Corollary}{Corollaries}
\newtheorem{lem}[thm]{Lemma}
\crefname{lem}{Lemma}{Lemmas}
\newtheorem{prop}[thm]{Proposition}
\crefname{prop}{Proposition}{Propositions}

\crefname{conj}{Conjecture}{Conjectures}

\crefname{ques}{Question}{Questions}
\theoremstyle{definition}

\crefname{defn}{Definition}{Definitions}
\newtheorem{rem}[thm]{Remark}
\crefname{rem}{Remark}{Remarks}

\crefname{ex}{Example}{Examples}

\crefname{obs}{Observation}{Observations}

\crefname{claim}{Claim}{Claims}

\crefname{ass}{Assumption}{Assumptions}

\numberwithin{thm}{section}

\newcommand{\cD}{\ensuremath{\mathcal D}}

\newcommand{\cL}{\ensuremath{\mathcal L}}

\newcommand{\cU}{\ensuremath{\mathcal U}}

\newcommand{\bbE}{{\ensuremath{\mathbb E}} }

\newcommand{\bbP}{{\ensuremath{\mathbb P}} }

\newcommand{\bbR}{{\ensuremath{\mathbb R}} }

\newcommand{\bbZ}{{\ensuremath{\mathbb Z}} }

\let\oldd\d
\renewcommand{\d}{{\ensuremath{\delta}}}

\newcommand{\h}{{\ensuremath{\eta}}}
\let\oldk\k
\renewcommand{\k}{{\ensuremath{\kappa}}}

\let\oldl\l
\renewcommand{\l}{{\ensuremath{\lambda}}}
\let\oldL\L
\renewcommand{\L}{{\ensuremath{\Lambda}}}
\newcommand{\m}{{\ensuremath{\mu}}}

\let\oldo\o
\renewcommand{\o}{{\ensuremath{\omega}}}
\let\oldO\O
\renewcommand{\O}{{\ensuremath{\Omega}}}
\newcommand{\p}{{\ensuremath{\pi}}}
\let\oldr\r
\renewcommand{\r}{{\ensuremath{\rho}}}

\let\oldt\t
\renewcommand{\t}{{\ensuremath{\tau}}}
\let\oldu\u
\renewcommand{\u}{{\ensuremath{\upsilon}}}

\renewcommand{\leq}{\leqslant}

\renewcommand{\le}{\leqslant}
\renewcommand{\ge}{\geqslant}
\renewcommand{\to}{\rightarrow}

\DeclareDocumentCommand \cD { o o } {%
  \IfNoValueTF {#1} {{\ensuremath{\mathcal{D}}} }%
  {\IfNoValueTF {#2} {{\ensuremath{\mathcal{D}^{\mathrm{\scriptstyle{#1}}}}}}%
  {{\ensuremath{\mathcal{D}^{\mathrm{\scriptstyle{#1}}}_{#2}}}}}}
\DeclareDocumentCommand \trel { o } {%
  \IfNoValueTF {#1} {{\ensuremath{T_{\rm rel}}} }{{\ensuremath{T_{\rm rel}^{\mathrm{\scriptstyle{#1}}}}}}%
}

\newcommand{\1}{{\ensuremath{\mathbbm{1}}} }
\newcommand{\tbp}{{\ensuremath{\tau_0^{\mathrm{BP}}}} }
\newcommand{\Tbp}{{\ensuremath{\m\left(\tbp\right)
}} }
\newcommand{\Et}{\ensuremath{\bbE_{\m}(\t_0)}}

\newcommand{\var}{\operatorname{Var}}

\newcommand{\bzero}{\ensuremath{\mathbf{0}}}

\newcommand{\qc}{q_{\mathrm{c}}}
\newcommand{\lgs}{\operatorname{long}}

\newcommand{\md}{\mathrm{\;\!d}}
\newcommand{\rd}{\r_{\mathrm{D}}}

\begin{document}
\title{Fredrickson--Andersen model in two dimensions}

\thanks{This work is supported by ERC Starting Grant 680275 ``MALIG.''}
\author{Ivailo Hartarsky}\address{CEREMADE, CNRS, UMR 7534, Universit\'e Paris-Dauphine, PSL University, Place du Mar\'echal de Lattre de Tassigny, 75016 Paris, France}

\author{Fabio Martinelli}\address{Dipartimento di Matematica e Fisica, Universit\`a Roma Tre, Largo S.L. Murialdo, 00146, Roma, Italy}

\author{Cristina Toninelli}\address{CEREMADE, CNRS, UMR 7534, Universit\'e Paris-Dauphine, PSL University, Place du Mar\'echal de Lattre de Tassigny, 75016 Paris, France}

\begin{abstract}
The present expository article overviews recent mathematical advances on the Fredrickson--Andersen kinetically constrained spin model in two dimensions. It was introduced in physics as a toy model for recovering the glassy phenomenology in supercooled liquids close to the glass transition via dynamic constraints as opposed to static interactions.
\end{abstract}

\begin{resume} Dans cet article expositoire on discute des avanc\'ees math\'ematiques r\'ecentes sur le mod\`ele cin\'etiquement contraint de  Fredrickson--Andersen en deux dimensions. Il fut introduit en physique comme mod\`ele jouet pour reproduire la ph\'enom\'enologie des liquides surfondus pr\`es de la transition vitreuse, moyennant une dynamique contrainte plut\^ot que des interactions statiques.
\end{resume}

\maketitle

\section{Models}
The Fredrickson--Andersen model (FA) is a family of models named after their inventors \cite{Fredrickson84}. They are paradigmatic examples within the larger class of \emph{kinetically constrained models} (see \cites{Toninelli21,Ritort03}). Their purpose is to provide an accessible toy model featuring glassy behaviour only via dynamical facilitation (see \cite{Speck19}), while having a trivial stationary state. From the mathematical viewpoint these interacting particle systems (see \cite{Liggett05}) are challenging to analyse as compared to close relatives such as the stochastic Ising model (see \cite{Martinelli99}) notably due to their lack of attractiveness. FA and, more generally, KCM have deep links to synchronous deterministic monotone processes known as bootstrap percolation (see \cite{Morris17a}), which we will be led to discuss in detail. Building on bootstrap percolation knowledge and developing new tools for tackling the more intricate FA dynamics, significant rigorous progress has been made, often settling controversial nonrigorous predictions. Our goal is to account for the current state of the art, primarily focusing on the most recent advances \cites{Hartarsky20FA,Hartarsky20CBSEP,Shapira20} in addition to earlier results \cites{Cancrini08,Martinelli19,Blondel13,Mountford19,Pillai17,Pillai19}. Although most techniques carry over to higher dimensions, we focus on the two-dimensional setting for the sake of simplicity and concreteness.
\subsection{Bootstrap percolation}
\label{sec:BP}
Let us start by introducing the $j$-neighbour bootstrap percolation model. Let $\O=\{0,1\}^{\bbZ^2}$ and call a site $x\in\bbZ^2$ \emph{infected} (or \emph{empty}) for $\o\in \O$ if $\o_x=0$ and \emph{healthy} (or \emph{filled}) otherwise. For fixed $0<q<1$, we denote by $\m_q$ the product Bernoulli probability measure with parameter  $1-q$  under which each site is infected with probability $q$. When confusion does not arise, we write $\mu=\mu_q$. Given $j\in\{1,2,3,4\}$, the $j$-neighbour bootstrap percolation model on $\bbZ^2$ is the monotone cellular automaton evolving as follows. Let $A_0\subset\bbZ^2$ be the set of \emph{initially infected sites} distributed according to $\m$. Then for any integer \emph{time} $t\ge 0$ we recursively define
\[
A_{t+1}=A_t\cup\left\{x\in\bbZ^2:|N_x\cap A_t|\ge 
j\right\},
\]
where $N_x$ denotes the set of nearest neighbours of $x$ in the usual graph structure of $\bbZ^2$. In other words, a site becomes infected \emph{forever} as soon as its constraint becomes satisfied, namely as soon as it has at least $j$ already infected neighbours. 

We denote by $[A]=\bigcup_{t\ge 0}A_t$ the \emph{closure} of $A\subset\bbZ^2$ and define the \emph{critical probability}
\begin{equation}
\label{eq:def:qc}\qc=\inf\left\{q\in[0,1]:\m_q\left([A]=\bbZ^d\right)>0\right\}.
\end{equation}
Another key quantity  for bootstrap percolation is the \emph{infection time of the origin} defined as $\tbp=\inf\{t\ge0:0\in A_t\}$, so that $\tbp<\infty$ a.s.\ for $q<\qc$.

\subsection{The Fredrickson--Andersen model}
\label{subsec:models}
We next introduce the FA model, a natural stochastic counterpart of bootstrap percolation and our main focus. 

For integers $1\le j\le 4$, the \emph{Fredrickson--Andersen $j$-spin facilitated model} (FA-$j$f) is the continuous time Markov process with state space $\O=\{0,1\}^{\bbZ^2}$ constructed as follows. Each site is endowed with an independent Poisson clock with rate $1$. At each clock ring the state of the site is updated to a  Bernoulli random variable with parameter $1-q$ subject to the crucial constraint that if the site has fewer than $j$ infected (nearest) neighbours currently, then the update is rejected. We refer to updates occurring at sites with at least $j$ infected neighbours at the time of the update as \emph{legal}.
\begin{rem}
Contrary to bootstrap percolation, the FA-$j$f process is clearly non-monotone  because of the possibile recovery of infected sites with at least $j$ infected neighbours. This feature is one of the major obstacles in the analysis of the process, along with the lack of attractiveness caused by the fact that more infections may make a healing update legal.
\end{rem}
It is standard to show (see \cite{Liggett05}) that the FA-$j$f process is well defined and it is reversible w.r.t.\ $\m_q$. For any function $f$ depending on finitely many spins, its Dirichlet form reads
\begin{equation}
    \label{eq:def:Dir}
    \cD(f)=\sum_{x\in\bbZ^d}\m\left(\1_{|\{y\in N_x:\h_y=0\}|\ge j}\cdot\var_x(f(\h))\right),
\end{equation}
where the average $\m$ is over $\h\in\O$, $\h_y$ is the state of the spin at $y$ and $\var_x$ stands for the average over $\h_x$, given the restriction of the configuration $\h$ to $\bbZ^2\setminus\{x\}$.
This enables the definition of a key charachteristic timescale---the \emph{spectral gap} or inverse \emph{relaxation time}
\begin{equation}
\label{eq:def:trel}
(\trel)^{-1}=\inf_{f\not\equiv\text{const}}\frac{\cD(f)}{\var(f)}.
\end{equation}
Alternatively, like for bootstrap percolation, one may prefer to investigate the first infection time of the origin under the FA-$j$f dynamics
\[
\t_0=\inf\{t\ge 0:\h_0(t)=0\}.
\]
A central goal is to precisely quantify the asymptotics as $q\to 0$ of $\trel$ and $\t_0$ for the stationary process with initial law $\m_q$. In fact, whenever possible, one would also like to treat non-equilibrium initial conditions, but this is mostly open, so unless otherwise stated, we consider the stationary process. Since it is clear that FA-$3$f and FA-$4$f have $\trel=\infty$ and $\t_0=\infty$ with positive probability for any $q<1$, we focus on $j\in\{1,2\}$.

\section{\mbox{FA-1f}}
Initial results on \mbox{FA-1f} were obtained in \cite{Cancrini08}, focusing on the relaxation time. There it was proved that $\trel$ is asymptotically $q^{-2}$ up to a logarithmic factor, more precisely there exists a constant $C$ s.t. for any $q\in(0,1)$ it holds
\begin{equation}
\label{eq:FA1f}C/q^2\leq \trel \leq \log(1/q)/(Cq^2)\end{equation}
As we will see, this reflects the fact that, to first order, as $q\to 0$ infections are typically isolated and perform a random walks, jumping at rate of order $1/q$ to neighbouring positions. In order to equilibrate, these random walks need to cover a volume of the order of their inverse density, namely $1/q$, hence the intuition behind the $1/q^2$ scaling.

More rigorously, the lower bound of \cite{Shapira20} consists in examining the number of connected clusters of infections truncated at distance $1/q$ from the origin as test function $f$ in \cref{eq:def:trel}. Since infections are rare, they are mostly isolated and $\var(f)$ scales like $q\cdot q^{-2}$. Moreover, the number of clusters changes by at most $4$ after a flip and only changes if the flip occurs at a site with two or more adjacent infections. Thus, contributions to the Dirichlet form of \cref{eq:def:Dir} only come from transitions with three infections at or around a given vertex, yielding $\cD(f)\approx q^{3}\cdot q^{-2}$. Hence, \cref{eq:def:trel} gives $\trel\ge 1/q^2$. A slightly more involved argument allows one to also deal with the expectation $\Et$ of $\t_0$ under the stationary process \cite{Shapira19}.

Rather than explaining the upper bound's original proof from \cite{Cancrini08}, we will take a simpler but less direct route by defining a closely related model of \emph{coalescing and branching simple exclusion process} (CBSEP) and then deducing the result on \mbox{FA-1f}. Essentially, \mbox{FA-1f} is CBSEP's evil twin lacking nice properties, but behaving exactly the same way.

\subsection{An auxiliary model: CBSEP}
\label{subsubsec:CBSEP}
Let $G=(V,E)$ be a connected graph. Minimum, maximum, and
average degrees in $G$ are denoted by $d_{\rm min}, d_{\rm max}$ and
$d_{\rm avg}$, respectively. The degree of $x \in V$ is denoted by $d_x$. For any $\o\in\O=\{0,1\}^V$ and any vertex $x\in V$ we say that $x$ is \emph{filled} (resp.\ \emph{empty}), or that there is a \emph{particle} (resp.\ \emph{hole}) at $x$, if $\o_x=1$ (resp.\ 0). We define $\O_+=\O\setminus \{\bzero\}$ to be the event that there exists at least one particle. Similarly, for any edge $e=\{x,y\}\in E$ we refer to $(\o_x,\o_y)\in \{0,1\}^{\{x,y\}}$ as the state of $e$ in $\o$ and write $E_e=\{\o\in\O:\o_x+\o_y\neq 0\}$ for the event that $e$ is not empty (its vertices are not both empty).

Given $p\in
(0,1)$, let $\pi=\bigotimes_{x\in V}\pi_x$ be the product Bernoulli measure in which each vertex is filled with probability $p$ and let $\mu(\cdot):=\pi(\cdot| \O_+)$ (if $G$ is infinite, then simply $\m=\p$). Given an edge
$e=\{x,y\}$, we write
$\pi_e:=\pi_x\otimes\pi_y$ and $\l(p):=\pi(E_e)=p(2-p)$.

\emph{CBSEP} is a continuous time Markov chain on $\O_+$ for which the state of any edge $e\in E$ such that $E_e$ occurs  is resampled with rate one w.r.t.\ $\pi_e(\cdot| E_e)$. Thus, any edge containing exactly one particle moves the particle to the opposite endpoint (the \emph{SEP move}) with rate $(1-p)/(2-p)$ and creates an extra particle at its empty endpoint
(the \emph{branching move})
with rate $p/(2-p)$. Moreover, any edge containing two
particles kills one of the two particles chosen uniformly (the \emph{coalescing move}) with rate $2(1-p)/(2-p)$. The chain is readily seen to be reversible w.r.t.\ $\mu$ and
ergodic on $\O_+$, because from any configuration we can reach the configuration with a
particle at each vertex.
If $c(\o,\o')$ denotes the jump rate from $\o$ to $\o'$, the Dirichlet
form $\cD[CBSEP](f)$ of the chain has the expression
\begin{equation}
\label{eq:def:cD:CBSEP}
\cD[CBSEP](f)=\frac 12 \sum_{\o,\o'}\mu(\o)c(\o,\o')\left(f(\o')-f(\o)
  \right)^2=\sum_{e\in E}\mu(\1_{E_e}\var_{e}(f| E_e)).\end{equation}

Notice that the branching and coalescing moves of {CBSEP} are exactly the moves allowed in \mbox{FA-$1$f}, if we identify the particles of CBSEP with the infected sites of FA-$1$f. Moreover, the SEP move for the edge $\{x,y\}$ from $(1,0)$ to $(0,1)$ can be reconstructed using two consecutive \mbox{FA-$1$f} moves, the first one filling the hole at $y$ and the second one emptying $x$. If we also take into account the rate for each move, we easily get the following comparison between the respective Dirichlet forms (see \cref{eq:def:Dir} and e.g.\ \cite{Levin09}*{Chapter 13.4}): there exists an absolute constant $c>0$ such that for all $f:\Omega_+\to \bbR$ it holds that
  \begin{equation}
    \label{eq:13}
    c^{-1}\cD[FA-1f](f)\le
    \cD[CBSEP](f)\le c d_{\rm max}p^{-1}  \cD[FA-1f](f),
  \end{equation}
setting the parameter $q$ of \mbox{FA-$1$f} equal to the parameter $p$ of CBSEP. In our application to \mbox{FA-$1$f} for $p\to0$ only the upper bound, which we believe to be sharper, counts.

Although the two models are clearly closely related, we would like to
emphasise that {CBSEP} has many advantages over \mbox{FA-$1$f}, making its study
simpler. Most notably, {CBSEP} is \emph{attractive} in
the sense that there exists a grand-coupling (see e.g.\ \cite{Levin09}) which preserves the partial order on $\O$
given by $\o\prec \o'$ iff $\o_x\le \o'_x$ for all $x\in V$. Furthermore, it is also natural to embed in {CBSEP} a continuous time random walk $(W_t)_{t\ge 0}$ on $G$ such that {CBSEP} has a particle at $W_t$ for all $t\ge 0$. The latter is a particularly fruitful feature, which is challenging to reproduce for \mbox{FA-$1$f} \cite{Blondel13}.

Thanks to \cref{eq:13,eq:def:trel}, in order to upper bound $\trel[FA-1f]$ and recover the result of \cite{Cancrini08}, it suffices to prove the following.
\begin{prop}
\label{prop:trel:CBSEP}
If $G=\bbZ^2$, then $\trel[CBSEP]\le O(\log(1/p)/p)$.
\end{prop}
A proof was given in \cite{Hartarsky20CBSEP} and \cite{Hartarsky20FA}*{Appendix B} up to minor modifications. In fact, much more is proved there. Namely, CBSEP on arbitrary graphs is treated, establishing often sharp bounds on $\trel$, but, more importantly, also on its logarithmic Sobolev constant.\footnote{This constant is defined like the spectral gap in \cref{eq:def:trel} with $\var(f)$ replaced by the entropy $\m(f^2\log(f^2/\m(f^2)))$.} A corollary of such stronger results and \cref{eq:13} is control of the mixing and \mbox{$L^2$-mixing} times of \mbox{FA-1f}. This recovers, strengthens and generalises results of Pillai and Smith \cites{Pillai17,Pillai19} proved in a different and somewhat more involved way.

In addition, \cites{Hartarsky20CBSEP,Hartarsky20FA} study a generalised version of CBSEP with general state spaces per site instead of $\{0,1\}$. For this generalised model they establish appropriate mixing time bounds crucial for the results on FA-2f discussed in \cref{subsec:FA2f}.

\section{\mbox{2-neighbour} bootstrap percolation}
\label{subsec:2n}
We next turn our attention to \mbox{$2$-neighbour} bootstrap percolation in two dimensions, which is a prerequisite for \mbox{FA-$2$f}. The \mbox{2-neighbour} bootstrap percolation originates from \cite{Chalupa79} (see also \cites{Pollak75,Kogut81}). Initially it was believed that $\qc>0$ based on simulations (see \cite{Adler88} and references therein) with estimated values in $(0.035,0.17)$. However, it was proved soon after \cite{VanEnter87} that in fact $\qc=0$. This was the first manifestation of what would grow to be called the \emph{bootstrap percolation paradox} we will keep returning to. To give it in a somewhat simplistic sentence, it refers to the observation that predictions on bootstrap percolation based on simulations always fail, no matter how advanced rigorous results they take into account. An early discussion of this paradox concerning the above can be found in \cite{VanEnter90}, while subsequent reassessments include \cites{DeGregorio04,Gravner08}.

\subsection{Coarse threshold}
The first quantitative statement in the domain of bootstrap percolation, which naturally laid its foundations is due to Aizenman and Lebowitz \cite{Aizenman88} (for nonrigorous precursors see \cite{Lenormand84}). They proved that
\begin{equation}
\label{eq:AL}
\tbp=\exp(\Theta(1/q))
\end{equation}
w.h.p. We provide a sketch of the argument, as it introduces ingredients essential to us. The first thing to note about \mbox{$2$-neighbour} bootstrap percolation is that the closure of any set of infections is the smallest (in terms of inclusion) collection of rectangles (with sides parallel to the axes of the lattice) at graph distance at least $3$ from each other containing the infections. Thus, the closure of any set can be determined via the following \emph{rectangles process}. We start off with a collection of rectangles consisting of each of the initial infections. At each step we merge two of them at graph distance $2$ or less, replacing them by the smallest rectangle containing their union. Repeating this until the process becomes stationary yields the collection of rectangles in the closure. A corollary of this process is the following fundamental lemma.
\begin{lem}[Aizenman--Lebowitz \cite{Aizenman88}]
\label{lem:AL}
We say that a rectangle $R$ is \emph{internally filled} (by the set $A$ of initial infections), if $[A\cap R]=R$. If $R$ is internally filled, then for every $k\le\lgs(R)$ there exists an internally filled rectangle $S\subset R$ such that $k\le \lgs(S)\le 2k$, where $\lgs(R)$ denotes the number of sites on the longer side of $R$.
\end{lem}
Clearly, $\tbp<\exp(c/q)$ implies that the origin belongs to an internally filled rectangle with long side at most $\exp(c/q)$ with $c$ to be chosen appropriately later. Then \cref{lem:AL} shows that within distance $\exp(c/q)$ of the origin there should be an internally filled rectangle $R$ of long side of our choice up to a factor $2$. The right side length to choose, which we refer to as \emph{critical scale}, is $1/q$. Observing that such an internally filled rectangle cannot contain two consecutive healthy rows/columns, we get
\[\m([A\cap R]=R)\le \left(1-\left(1-q\right)^{2\lgs(R)}\right)^{\lfloor\lgs(R)/2\rfloor}=\exp\left(-\Theta(1/q)\right),\]
concluding the proof that $\tbp\ge \exp(\O(1/q))$ w.h.p.\ by the union bound on all possible positions of $R$, choosing $c$ small enough.

A matching upper bound is guided by a similar idea (explaining the title `Metastability effects in bootstrap percolation' of \cite{Aizenman88}). We first make sure to internally fill a square of (supercritical) side, say, $q^{-3}$ and then this \emph{critical droplet} is likely to grow and infect the entire grid at roughly linear speed. The internal filling can be directly forced starting from one infection and asking for it to find another one on its right and top side on each line as it progressively infects a growing square. This has probability
\begin{equation}
\label{eq:AL:computation}
q\prod_{k=1}^\infty\left(1-\left(1-q\right)^{k}\right)^2\approx\exp\left(2\int_0^{\infty}\log\left(1-e^{-qx}\right)\md x\right)=\exp\left(-\Theta(1/q)\right)\end{equation}
and thus is likely to occur within distance $\exp(C/q)$ of the origin for $C$ large enough. We may then ensure that with overwhelming probability every vertical or horizontal line of length $q^{-3}$ at distance at most $\exp(C/q)$ from the origin contains an infection, so that the critical droplet does grow roughly linearly until it engulfs the origin after time $\exp(O(1/q))$, proving \cref{eq:AL}.

\subsection{Sharp threshold}
Naturally, following \cref{eq:AL} the question of the day became determining the implicit constant. This came about in a breakthrough of Holroyd 15 years later \cite{Holroyd03}, proving that w.h.p.
\begin{equation}
\label{eq:Holroyd}\tbp=\exp\left(\frac{\pi^2+o(1)}{18q}\right).\end{equation}
We will prove stronger lower and upper bounds in the sequel, so it is useful to give an idea of the proof, which introduced several crucial techniques commonly used thereafter. As in the Aizenman--Lebowitz result, the main difficulty is controlling the probability of a rectangle of size roughly $1/q$ being internally filled. More precisely, \cref{eq:Holroyd} follows once we show that 
for $R$ of size $C/q$ for $C$ large
\begin{equation}
\label{eq:BP:droplet:proba}
\m([A\cap R]=R)\approx e^{-\pi^2/(9q)}.
\end{equation}

\begin{figure}
  \centering  
  \begin{tikzpicture}[>=latex,x=0.5cm,y=0.5cm]
\draw[fill] (8,1) rectangle (11,4);
\draw (8,1) rectangle (16,9);
\draw (11,4) -- (13,4);
\draw (13,1)--(13,4);
\draw[->] (11,2.5)--(13,2.5);
\draw (8,6)--(13,6);
\draw (13,6)--(13,1);
\draw[->] (10.5,4)--(10.5,6);
\draw (13,6)--(16,6);
\draw[->] (13,3.5)--(16,3.5);
\draw[->] (12,6)--(12,9);
\end{tikzpicture}
\qquad
\begin{tikzpicture}[>=latex,x=0.5cm,y=0.5cm]
\draw[fill] (10,3.3) rectangle (13,6.3);
\draw  (8, 1)-- (8,9);
\draw  (16,1)-- (16,9);
\draw  (8, 1)-- (16,1);
\draw  (8, 9)-- (16,9);
\draw  (8, 7.5)-- (16,7.5);
\draw  (8, 2.5)-- (16,2.5);
\draw [->] (12,7.5) -- (12,9);
\draw [->] (11,6.3) -- (11,7.5);
\draw [->] (11,3.3) -- (11,2.5);
\draw [->] (12,2.5) -- (12,1);
\draw [->] (14,4) -- (16,4);
\draw [->] (9,4) -- (8,4);
\draw [->] (10,5) -- (9,5);
\draw [->] (13,5) -- (14,5);
\draw  (9, 2.5)-- (9,7.5);
\draw  (14, 2.5)-- (14,7.5);
\draw  (9, 3.3)-- (14,3.3);
\draw  (9, 6.3)-- (14,6.3);
\draw  (10, 3.3)-- (10,6.3);
\draw  (14, 3.3)-- (14,6.3);
\end{tikzpicture}
\caption{Recursive structure of droplets used in 2-neighbour bootstrap percolation \cite{Holroyd03} and FA-2f \cite{Hartarsky20FA} respectively. Arrows indicate regions with no two consecutive healthy lines.}
\label{fig:super-good}
\end{figure}
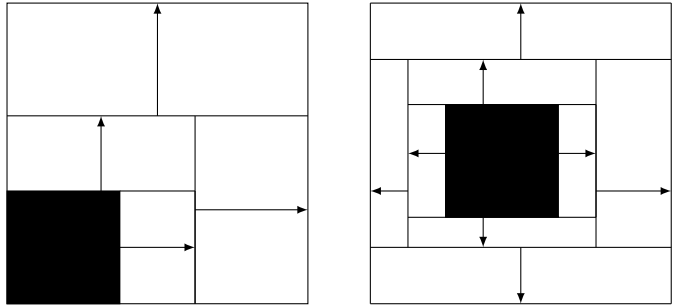

We will only discuss the easier lower bound in \cref{eq:BP:droplet:proba}. Once again we start from a single infection and make it infect progressively larger rectangles. However, it grows by an amount larger than $1$ in each direction before switching to the other (see \cref{fig:super-good}). Namely, the right choice is to grow in steps of $1/\sqrt{q}$. The use of this is that we do not need an infection on every line, but on every second line. This is the origin of the constant $\pi^2/9$: it arises like the integral in \cref{eq:AL:computation}, but for a function corresponding to the lack of two consecutive rows/columns of healthy sites. If one thinks about the two-term recurrence relation this function should come from (we only need to remember if an infection was found on the previous line or the one before it), it is not surprising that it appears as the root of a certain quadratic equation. The reader interested in the links of this function and its integral with integer partitions may consult \cites{Holroyd03a,Bringmann12}. Actually, the sketch above is not quite the way the result is proved in \cite{Holroyd03}, but anticipates \cite{Gravner08} and \cite{Hartarsky20FA} discussed below.

\subsection{Speed of convergence}
\Cref{eq:Holroyd} might as well have been the end of the story, had it not been a new manifestation of the bootstrap percolation paradox. Numerical estimates \cites{Adler89,Adler91,Nakanishi86} of the constant $\pi^2/18$ above had yielded less than half the correct value. This naturally leads to the question of how fast the convergence in \cref{eq:Holroyd} is. For this reason, we quantify the error term in \cref{eq:Holroyd}, again contradicting simulation predictions \cite{Teomy14} (see also \cite{Gravner12} for more) and showing that the convergence is very slow.
\begin{thm}[Second term]
\label{th:2n}
For \mbox{2-neighbour} bootstrap percolation in two dimensions it holds w.h.p.
\[\tbp=\exp\left(\frac{\pi^2-\Theta(\sqrt{q})}{18q}\right).\]
\end{thm}
The upper bound was established in \cite{Gravner08} and is based on the mechanism presented for \cref{eq:Holroyd}. Roughly speaking, the main difference, which is at the origin of the negative sign of the second term, is taking entropy into account. More precisely, rather than growing our squares in steps of $1/\sqrt{q}$, we allow the exact length of these increments to vary, while being of order $1/\sqrt{q}$. The entropy gained from this is sufficient to outweigh the energetic cost of deviating from a square shape.

The lower bound is significantly harder and is the subject of \cite{Hartarsky19}.

\section{\mbox{FA-2f}}
\label{subsec:FA2f}
Moving on to \mbox{FA-2f} (again in two dimensions), the story is much shorter. Indeed, the analogues of all the results for \mbox{2-neighbour} bootstrap---from the 1988 Aizenman--Lebowitz \cite{Aizenman88} one (\cref{eq:AL}) to the recent \cref{th:2n}---were not known before the recent contribution \cite{Hartarsky20FA}. Our task is then to review the only two previous rigorous results \cites{Cancrini08,Martinelli19} and copious nonrigorous ones.

\subsection{Background}
As for bootstrap percolation, the initial expectation was that \mbox{FA-2f} would exhibit a nontrivial transition \cite{Fredrickson84}. We should emphasise that here and in the other works to be quoted below, predictions were made, taking into account bootstrap percolation results already available. In particular, a transition was expected despite its absence in bootstrap percolation \cite{Fredrickson85}. This was quickly dissipated by physicists \cites{Fredrickson86,Reiter91}, though rigorous results in this direction came only two decades later \cite{Cancrini08} (see also \cite{Cancrini07}). Denoting the semigroup of FA-2f by $(P_t)_{t\ge 0}$, the \emph{ergodicity critical parameter} is defined as
\[\qc=\inf\left\{q>0:\forall f\in L^2(\m),\lim_{t\to\infty}P_t f=\m(f)\right\}.\]
It was proved in \cite{Cancrini08} that this transition coincides with the one of 2-neighbour bootstrap percolation (\cref{eq:def:qc}), which is why we still denote it $\qc$. It also coincides with the more standard ergodic theory definition: for $q>\qc$ the eigenvalue $0$ of $\cL_\cU$ is simple and, therefore, by the ergodic theorem we also have
\[\qc=\inf\{q>0:\bbP_\m(\t_0<\infty)=1\}.\]

The same paper also discarded the possibility that for \mbox{FA-2f} e.g.\ the tail $\bbP_\m(\t_0\ge t)$ of the infection time would decay as a stretched exponential.\footnote{In \cite{Cancrini08}, the first time when the origin changes state was considered, rather than the time when it becomes infected, but this is unimportant.} The pure exponential decay they established was quite unexpected as numerous nonrigorous works had exhibited evidence of stretching, though with various stretching exponents \cites{Fredrickson85,Reiter91,Fredrickson86, Graham93,Graham97,Harrowell93,Fredrickson88, Butler91,Alers87,Fredrickson86a} according to \cites{Cancrini07,Ritort03}. The exponential decay of the above quantity follows rather easily, once it is established that $\trel<\infty$, though this had seemingly eluded physicists, who also had various predictions for the scaling of $\trel$ as $q\to 0$, as we will see.

The last results of \cite{Cancrini08} for \mbox{FA-2f} are the quantitative bounds on $\trel$
\begin{equation}
\label{eq:FA2f:CMRT}\exp\left(\frac{\pi^2-o(1)}{18q}\right)=\O\left(\Tbp\right)\le \Et\le \trel/q\le \exp\left(O\left(1/q^5\right)\right),
\end{equation}
in particular establishing that it is finite. The first two inequalities hold in great generality and are not hard. The upper bound is both harder and not useful to us, so we do not discuss it further. Unfortunately, \cref{eq:FA2f:CMRT} does not give the precise scaling of $\Et$. Therefore, discriminating between the conflicting expressions suggested by physicists \cites{Reiter91,Fredrickson88,Butler91, Fredrickson86,Nakanishi86,Toninelli05,Teomy15, Graham93,Graham97} remained an open problem (e.g.\ \cite{Ritort03} asked for settling this controversy). Progress in this direction was made recently in \cite{Martinelli19}, improving the upper bound to
\begin{equation}\label{eq:FA2f:old}
\exp\left(\frac{O(\log^2(1/q))}{q}\right),
\end{equation}
much closer to the lower one, but still inconclusive. Indeed, by 2019, when \cite{Martinelli19} was published, several (different) predictions not only for the presence or absence of a logarithmic factor but also on the potential sharp constant, based on \cref{eq:Holroyd}, had been accumulated in 35 years. The proof of \cite{Martinelli19} is again not very useful to us, so we do not discuss it.

Before settling the matter, let us explain the different predictions. The first one appeared in \cite{Nakanishi86}, where, based on numerical simulations, a faster than exponential divergence in $1/q$ was conjectured. The first to claim an exponential scaling $\exp(\Theta(1)/q)$ was Reiter \cite{Reiter91}. He argued that the infection process of the origin is dominated by the motion of `macro-defects,' i.e.\ rare regions having probability $\exp(-\Theta(1)/q)$ and size $q^{-\Theta(1)}$ that move at an exponentially small rate $\exp(-\Theta(1)/q)$. Later \cite{Toninelli05} refined the above picture. There it was argued that macro-defects should coincide with the critical droplets of $2$-neighbour bootstrap percolation, having probability $\exp(-\pi^2/(9 q))$ and that  
the time scale of the relaxation process inside a macro-defect should be $\exp(c/\sqrt q)$, i.e.\ sub-dominant with respect to the inverse of their density, in sharp contrast with the prediction of \cite{Reiter91}. Based on this and on the idea  that macro-defects move diffusively, the relaxation time scale of \mbox{FA-2f} in $d=2$ was conjectured to diverge as $\exp(\pi^2/(9q))$ \cite{Toninelli05}*{Section 6.3}. 
Yet, a different prediction was later made in \cite{Teomy15} implying a different scaling of the form $\exp(2\pi^2/(9q))$. 

\subsection{Result}
The main result of \cite{Hartarsky20FA} shows that the scaling prediction of \cites{Toninelli05,Reiter91} is correct, contrary to those of \cites{Teomy15,Nakanishi86}. Moreover, they show that the characteristic time scale of the relaxation process \emph{inside} a macro-defect agrees with the prediction of \cite{Toninelli05} and disproves the one of \cite{Reiter91}.
\begin{thm}
\label{th:FA2f}
As $q\to 0$ the stationary \mbox{FA-$2$f} model on $\bbZ^2$  satisfies: 
\begin{align}
\label{eq:FA2f:lower}
\Et&\ge{}\exp\left(\frac{\pi^2}{9q}\left(1-\sqrt{q}\cdot O(1)\right)\right),\\
\label{eq:FA2f:upper}\Et&\leq{}\exp\left( \frac{\pi^2}{9q}\left(1+\sqrt{q}\cdot\left(\log(1/q)\right)^{O(1)}\right)\right).
\end{align}
Moreover, these also hold for $\t_0$ w.h.p.
\end{thm}

\begin{rem} Despite the resemblance, \cref{th:FA2f} is by no means a corollary of \cref{th:2n}. While the lower bound \cref{eq:FA2f:lower} does indeed follow rather easily from \cref{th:2n} together with an improvement of the `automatic' lower bound $\Et\ge\O\left(\Tbp\right)$ from \cref{eq:FA2f:CMRT}, the proof of \cref{eq:FA2f:upper} is much more involved. In particular, it requires guessing an efficient infection/healing mechanism to infect the origin, which has no counterpart in the monotone \mbox{2-neighbour} bootstrap percolation model.
\end{rem}

\subsection{Behind Theorem \ref{th:FA2f}: high-level ideas}
\label{subsec:heuristics}
The main intuition behind \cref{th:FA2f} is that for $q\to 0$ the relaxation to equilibrium of the stationary  \mbox{FA-$2$f} process is dominated by the slow motion of unusually unlikely patches of infection, dubbed \emph{mobile droplets} or just \emph{droplets}. In analogy with the \emph{critical droplets} of bootstrap percolation, mobile droplets have a linear size which is polynomially increasing in $q$ (with some arbitrariness), i.e.\ they live on a much smaller scale than the metastable length scale $e^{\Theta(1/q)}$ arising in bootstrap percolation. One of the main requirements dictating the choice of the scale of mobile droplets is the requirement that the typical infection environment around a droplet is w.h.p.\ such that the droplet is able to move under the \mbox{FA-$2$f} dynamics in any direction. Within this scenario the main contribution to the infection time of the origin for the stationary \mbox{FA-$2$f} process should come from the time it takes for a droplet to reach the origin. 

In order to translate the above intuition into a mathematically rigorous proof, one is faced with two different fundamental problems: 
\begin{enumerate}
\item\label{item:definition} a precise, yet workable, definition of mobile droplets;
\item\label{item:evolution} an efficient model for their `effective' random evolution.
\end{enumerate}
In \cites{Martinelli19,Martinelli19a,Hartarsky21a} mobile droplets (dubbed `super-good' regions there) have been defined rather rigidly as fully infected regions of suitable shape and size and their motion has been modelled as a \emph{generalised \mbox{FA-$1$f} process} on $\bbZ$ \cite{Martinelli19a}*{Section 3.1}. In the latter process mobile droplets are freely created or destroyed with the correct heat-bath equilibrium rates but \emph{only at locations which are
adjacent to an already existing droplet}.

While rather powerful and robust, this solution has no chance to give the \emph{exact} asymptotics of either \ref{item:definition}, or \ref{item:evolution} above. Indeed, a mobile droplet should be allowed to deform itself and move to a nearby position like an amoeba, by rearranging its infection using the \mbox{FA-$2$f} moves. This `amoeba motion' between nearby locations should occur on a time scale much smaller than the global time scale necessary to bring a droplet from far away to the origin. In particular, it should not require to first create a new droplet from the initial one and only later destroy the original one (the main mechanism of the droplet dynamics under the generalised \mbox{FA-$1$f} process).  

The work \cite{Hartarsky20FA} offered a solution to \ref{item:definition} and \ref{item:evolution} above which indeed leads to determining the exact asymptotics of the infection time. Concerning \ref{item:definition}, their treatment consists of two steps. They first propose a sophisticated multiscale definition of mobile droplets which, in particular, introduces a  crucial degree of \emph{softness} in their microscopic infection's configuration. Namely, on each scale the lower scale droplet is allowed to have an arbitrary position within the higher scale one (see \cref{fig:super-good}), instead of systematically being at the bottom-left corner. This allows for the droplet to move by rearranging its inside, which corresponds to changing the position of its core on each scale.

The second and much more technically involved step is developing the tools necessary to analyse the \mbox{FA-2f} dynamics inside a mobile droplet. In particular, \cite{Hartarsky20FA} then proves two key features:
\begin{enumerate}[label=(1.\alph*),ref=(1.\alph*)]
\item\label{item:rd} to the leading order the probability $\rd$ of mobile droplets matches \cref{eq:BP:droplet:proba}:
\[\rd
\ge \exp\left(-\frac{\pi^2}{9q}-\frac{O(\log^2(1/q))}{\sqrt q}\right),\]
\item\label{item:gd} the `amoeba motion' of mobile droplets to
a nearby location occurs on time scale $\exp(O(\log(1/q)^3)/{\sqrt q})$ which is sub-leading w.r.t.\ the main time scale
of the problem and only manifests in the second term of \cref{eq:FA2f:upper}.
\end{enumerate}
Property \ref{item:rd} follows rather easily essentially as explained for \cref{eq:BP:droplet:proba}, while proving property \ref{item:gd} required several ideas. In particular, let us mention how the use of the standard technique of canonical paths is avoided. One proves a Poincar\'e inequality on each scale for the product measure $\m$ on the volume of the droplet, conditioned to have the droplet structure present. Deducing the higher-scale inequality from the lower scale one, on the other hand, relies on a suitable adaptation the bisection technique from \cite{Cancrini08}.

While properties \ref{item:rd} and \ref{item:gd} above are essential, they are not sufficient on their own for solving problem \ref{item:evolution} above. That is where CBSEP's key role is to be played (recall \cref{subsubsec:CBSEP}). In \cite{Hartarsky20FA} the random evolution of mobile droplets is treated as a generalised CBSEP at the level of Poincar\'e inequalities. More precisely, they considered a renormalisation, so that the state of a box is 1 if it contains a droplet and 0 otherwise. Thus, the parameter $p$ of CBSEP corresponds to the density $\rd$ of droplets. We next recall the relaxation time of CBSEP from \cref{prop:trel:CBSEP}, as well as the definitions \cref{eq:def:trel,eq:def:cD:CBSEP}. Finally, we observe that the terms $\var_e(f|E_e)$ in \cref{eq:def:cD:CBSEP} call out for a conditional Poincar\'e inequality for a droplet, as discussed in the context of \ref{item:gd}. Putting these ingredients together, one does obtain the scaling of \cref{th:FA2f}.

\section{Open problems}

The results we accounted for in the previous sections concern the behaviour of the stationary process started from the equilibrium measure $\mu$.
A key issue, both from the mathematical and physical points of view, is to analyse the out of equilibrium dynamics, namely the evolution starting from an initial measure $\nu\neq \mu$. 
For $\nu$ a product Bernoulli measure of parameter $1-q'$ with $q'\in[0,1)$, we expect convergence to equilibrium to occur for FA-1f and FA-2f. Namely  we expect 
$$\lim_{t\to \infty} \nu(P_t f(\eta))=\mu(f)$$ to hold 
for any local function $f$, where $P_t$ denotes the semigroup of the process. Progress in this direction has been hindered by the lack of robust tools. Indeed, due to non-attractiveness, many powerful tools that have been developed to analyse relaxation to equilibrium for other interacting particle systems, such as coupling and censoring arguments, are unavailable here. Furthermore, due to the presence of constraints, a worst case analysis is not possible (if the initial configuration is completely healthy, no site can ever be infected and equilibrium is never attained). Similarly, the Sobolev and logarithmic Sobolev constants diverge in infinite volume, discarding classic arguments for proving relaxation to equilibrium  based on hypercontractivity (see \cite{Holley87}). 

The only results available in this out of equilibrium setting concern FA-1f and establish convergence to equilibrium  only in the the restricted parameter region $q>1/2$ \cite{Blondel13}. As for the speed of convergence, the result of \cite{Blondel13} proves convergence at least as fast as a stretched exponential with exponent $1/2$. This was later improved to exponential convergence in \cite{Mountford19}, albeit in an even more restricted parameter region (for $q$ sufficiently close to 1). The reason why the above results hold only in a restricted region is that, in order to circumvent the problems stemming from non attractiveness, both \cites{Blondel13,Mountford19} compare FA-1f to an auxiliary dynamics. These dynamics have the advantage of being attractive, but the disadvantage of producing less infected sites and being efficient only if $q$ is sufficiently high. For FA-2f even the regime $q$ close to 1 is completely open due to the cooperative dynamics (no finite set of infections is able to move around on its own).

Let us conclude by mentioning two open questions for the stationary process started from $\mu$.
\begin{enumerate}[label=(\roman*),ref=(\roman*)]
    \item For FA-$1$f a logarithmic gap remains to be filled between the upper and lower bounds to obtain the scaling of $\trel$ (see \cref{eq:FA1f}).
    \item For FA-$2$f the  second order term in the exponent remains to be determined to reach the precision of \cref{th:2n} for bootstrap percolation or even determine its sign.
\end{enumerate}

\let\d\oldd
\let\k\oldk
\let\l\oldl
\let\L\oldL
\let\o\oldo
\let\O\oldO
\let\r\oldr
\let\t\oldt
\let\u\oldu

\bibliography{bib}
\end{document}